\documentclass[12pt,english]{article} \usepackage{graphicx} \usepackage{amssymb} \usepackage{amsmath} \usepackage{babel} \textwidth=6.5in \oddsidemargin=0in \topmargin=-.9in \textheight=9.5in 

\begin{document}

\title{\textbf{Classical Simulation of Quantum Fields I }}

\author{T. Hirayama\;\;and B. Holdom\thanks{bob.holdom@utoronto.ca}\\ \emph{\small Department of Physics, University of Toronto}\\[-1ex] \emph{\small Toronto ON Canada M5S1A7}} \date{} \maketitle \begin{abstract} We study classical field theories in a background field configuration where all modes of the theory are excited, matching the zero-point energy spectrum of quantum field theory. Our construction involves elements of a theory of classical electrodynamics by Wheeler-Feynman and the theory of stochastic electrodynamics of Boyer. The nonperturbative effects of interactions in these theories can be very efficiently studied on the lattice. In $\lambda\phi^{4}$ theory in $1+1$ dimensions we find results, in particular for mass renormalization and the critical coupling for symmetry breaking, that are in agreement with their quantum counterparts. We then study the perturbative expansion of the $n$-point Green's functions and find a loop expansion very similar to that of quantum field theory. When compared to the usual Feynman rules, we find some differences associated with particular combinations of internal lines going on-shell simultaneously. \end{abstract}

\section{Introduction}

Over the years various classical models have been proposed that are able to mimic at least some features of quantum mechanics. These attempts help to fuel debates on the interpretation of quantum mechanics. In this paper we will try to move the discussion from quantum mechanics toward quantum field theory, where the latter could be viewed as the more fundamental description of our quantum world. The content of a quantum field theory is said to reside in its $n$-point Green's functions, and so we are led to wonder just how close the $n$-point functions calculated in a classical context can come to the quantum counterparts.

We shall model the zero point fluctuations of quantum field theory by exciting all the modes of classical field theory in a corresponding way and averaging over their random phases. Their amplitudes are such that the energy per mode is $\frac{1}{2}\hbar\omega$. We may then wonder about the effects of interactions on such configurations when they are allowed to evolve via the full classical equations of motion. This leads us to develop numerical simulations of these fluctuations in an interacting classical theory, where the effects of interactions can be probed through the study of correlation functions. In particular we study $\lambda\phi^{4}$ theory in $1+1$ dimensions. We are able to obtain the value of the critical coupling for symmetry breaking by looking for signals of vanishing mass and the nonvanishing of an order parameter. The results are surprisingly consistent with those of the quantum field theory. A more detailed study of the lattice system, with an emphasis on mass renormalization, and its comparison to standard lattice methods applied to the quantum theory are presented in a companion paper \cite{C}. We achieve sufficient precision to extract a 2-loop effect, and we find a result in agreement, within errors, with the quantum loop calculation.

We note that there is a large body of work that uses the real time evolution of classical fields as an approximation scheme for the study of quantum field theories. We give a selection of such references in \cite{thermpert,therm,thermstable}. That work is concerned with thermal systems or other cases (such a inflaton decay and reheating in the early universe) where the particle numbers become large, in particular when the occupation numbers are large compared to the quantum vacuum ($n_k + 1/2\gg 1/2$). In such cases the classical dynamics dominates over quantum effects and the classical equations of motion may be used as an approximation, with an average over a classical ensemble of initial conditions weighted with the Boltzmann factor in the thermal case, for example. Although our numerical methods are similar they are not the same, since in our average over initial conditions we average only over the phases. We are also expressly comparing theories in the opposite regime, that of zero temperature. There is no a priori reason to expect that the interacting quantum vacuum state, where there are no classical excitations, to be closely simulated by any classical theory.

After we describe our lattice simulation we will then develop a perturbation theory directly in the classical theory that is expressly tailored to describe interactions in the classical background. The Lorentz invariance of the background, as implied by the $\frac{1}{2}\hbar\omega$ spectrum, is directly incorporated.  This development serves to make clear how renormalization effects arise in the classical theory, and from it we can make a direct comparison to quantum field theory at zero temperature. Perturbation theory has been used before, mainly in the context of the Schwinger-Keldysh formalism, to compare quantum and classical theories \cite{thermpert}.  Our conclusions are consistent with this previous work, namely that there are missing contributions in the diagrammatic expansion of the classical theory in comparison with the quantum theory.

Given that we are at zero temperature, where there is no reason to expect that the missing contributions are unimportant, we may then wonder why a discrepancy did not appear in our precision study of mass renormalization in \cite{C}. Moreover at strong coupling we may wonder why a classical simulation should even remotely resemble the quantum theory.  We now believe that the close resemblance that we do find depends crucially on the fact that phases and not amplitudes are being averaged over in our simulations. This is in contrast to previous simulations \cite{thermpert,therm,thermstable} where mode amplitudes are distributed according to a Gaussian distribution. The effect of fixed amplitudes or nongaussian distributions in the classical ensemble has been studied in more detail by one of the authors \cite{H}.

In the classical picture there is a quantity that corresponds to $\hbar$; this is not a fundamental quantity since it simply characterizes the overall normalization of the background solution. The question of the origin of these fluctuations and their amplitude leads to further speculative issues. In previous work \cite{E} we studied solutions to the gravitational field equations in which a negative cosmological constant drives rapid oscillations of the metric, as well as exciting the positive and negative energy modes of other fields. Due to the role of gravity we found stability of these solutions at the classical level.\footnote{This is despite the existence of negative energy modes, which in turn can help to regulate the total energy density.} The result is that spacetime can appear to be flat on large scales while still responding to the effects of a large cosmological constant. To be at all realistic the classical fields must exhibit these gravitationally induced fluctuations in a Lorentz invariant manner. Then these fluctuations could correspond to the classical fluctuations studied in this paper. The unusual aspect of this picture is that the cosmological constant is the fundamental quantity, and $\hbar$ is a derived quantity that follows from it. We mention this picture here because this is what led us to our present study.

In summary we study a collection of random classical fields, fields solving classical equations of motion, and demonstrate that their correlation functions resemble quite surprisingly the correlation functions of a quantum field theory. Our comparison involves both perturbative and strongly interacting systems at zero temperature, a comparison that has not been considered elsewhere in the literature. We are not claiming a priori that the classical system is a controlled approximation to quantum field theory, and we only consider the latter as a benchmark to which to compare our results from classical fields. But the similarities observed between the classical and quantum theories may have implications for our understanding of the foundations of quantum field theory \cite{H}.

In the next section we split the Feynman propagator into two parts and obtain a classical interpretation for each part. In section 3 we give a lattice formulation of the classical theory that both defines the theory and allows for its nonperturbative study. Section 4 presents some results from the lattice that show critical behavior at a coupling consistent with the quantum value. In sections 5 and 6 we develop a perturbation theory where we will see explicitly the emergence of loop effects. In section 7 we give simple rules for the construction of graphs and see how they deviate from quantum field theory. We focus on the 2-point function in section 8. In section 9 we make some initial remarks on fermions before concluding in section 10.

\section{The Feynman propagator}

Consider the Feynman propagator for a scalar field with mass $m$.\begin{eqnarray} D_{F}(x-x') & = & \hbar\int\frac{d^{4}p}{(2\pi)^{4}}\frac{i}{p^{2}-m^{2}+i\varepsilon}e^{- ip\cdot(x-x')}\nonumber \\ & = & \hbar\int\frac{d^{4}p}{(2\pi)^{4}}\left[P\left(\frac{i}{p^{2}-m^{2}}\right)+\pi\delta(p^{2}-m^{2})\right]e^{-ip\cdot(x-x')}\nonumber \\ & = & \hbar\int\frac{d^{3}p}{(2\pi)^{3}}\frac{1}{2\omega}\left[i\sin(\omega|t- t'|+\mathbf{p\mathnormal{\cdot}(x\mathnormal{-}x\mathnormal{'})})+\cos(\omega(t-t')+\mathbf{p\mathnormal{\cdot}(x\mathnormal{-}x\mathnormal{'})} )\right]\nonumber \\ & = & D_{P}(x-x')+D_{\delta}(x-x'),\qquad\omega=\sqrt{\mathbf{p}^{2}+m^{2}}\label{k}\end{eqnarray} The first term, the imaginary part of the Feynman propagator corresponding to the principal value part in the momentum representation, is a Green's function solution of\begin{equation} (\partial_{x}^{2}+m^{2})D_{P}(x-x')=-i\hbar\delta^{4}(x-x'),\label{f}\end{equation} corresponding to time symmetric, $t\leftrightarrow-t$, boundary conditions. This is the average of the advanced and retarded Green's functions. The real part of the Feynman propagator, the $\delta$-function piece, is a smooth function of coordinates and satisfies $(\partial_{x}^{2}+m^{2})D_{\delta}(x-x')=0$.

In quantum field theory the Feynman propagator is given as the expectation value of the time ordered product of quantum fields. This product, for a scalar field, can be written as\begin{equation} T\phi(x)\phi(x')=\frac{\mathrm{sgn}(t-t')}{2}[\phi(x),\phi(x')]+\frac{1} {2}\{\phi(x),\phi(x')\}.\end{equation} The first term is just a $c$-number and is precisely $D_{P}(x-x')$. The second term is a quantum operator whose expectation value is $D_{\delta}(x-x')$. This illustrates the classical nature of $D_{P}$ versus the quantum mechanical nature of $D_{\delta}$. Note that, peculiar as it seems, the quantum piece of the propagator has the on-shell $\delta$-function, while it is the classical piece $D_P$ that allows internal lines in momentum space Feynman rules to go off-shell.

A theory of electrodynamics based on the average of advanced and retarded Green's functions (a massless version of $D_{P}$), was shown by Wheeler and Feynman \cite{D} to nicely account for the otherwise ad hoc radiation damping term describing the motion of accelerating charges. They showed that their time symmetric formulation does not violate causality and is equivalent to the standard retarded Green's function description as long as all radiation is eventually absorbed. From the point of view of quantum field theory, a missing element of their theory is $D_{\delta}$.

In this paper we shall explore a picture for the origin of $D_{\delta}$ in terms of a fluctuating classical background configuration. A classical interpretation for $D_{\delta}$ is perhaps not completely unexpected, given the on-shell nature of $D_{\delta}$. The fluctuating background will be a superposition of all the plane-wave modes of the theory, of the form $\cos(\omega_\mathbf{p} t+\mathbf{p}\cdot\mathbf{x}+\theta_{\mathbf{p}})$, where each has a phase $\theta_{\mathbf{p}}$. If we denote the background configuration by $\phi_{0}(x)$, it can be obtained from the quantum field operator by replacing the annihilation and creation operators by phase factors, $a_{\mathbf{p}}\propto\sqrt{\hbar}e^{i\theta_{\mathbf{p}}}$ and $a_{\mathbf{p}}^{\dagger}\propto\sqrt{\hbar}e^{-i\theta_{\mathbf{p}}}$. This classical configuration has the zero-point energy spectrum of quantum field theory. In addition the following result emerges, involving an averaging over the phases of the modes denoted by $\langle\cdot\rangle_\theta$, \begin{equation} \left\langle \phi_{0}(x)\phi_{0}(x')\right\rangle _{\theta}=D_{\delta}(x-x').\label{bb}\end{equation} We stress that the $\hbar$ in $D_{\delta}$ arises as an overall normalization of classical modes.

We discuss the derivation of (\ref{bb}) in the next section where we use a finite volume lattice regularization. The basic ingredient in the derivation is\begin{equation} \int\prod_{\mathbf{k}}\left[\frac{d\theta_{\mathbf{k}}}{2\pi}\right]\sum_{\mathbf{p}}\cos(\omega_\mathbf{p} t+\mathbf{p}\cdot\mathbf{x}+\theta_{\mathbf{p}})\sum_{\mathbf{q}}\cos(\omega_\mathbf{q} t'+\mathbf{q}\cdot\mathbf{x}'+\theta_{\mathbf{q}})=\frac{1}{2}\sum_{\mathbf{p}}\cos(\omega_\mathbf{p}(t-t')+\mathbf{p}\cdot(\mathbf{x}-\mathbf{x}'))\label{j}\end{equation} for $\mathbf{k}$, $\mathbf{p}$, $\mathbf{q}$ in a discrete set of momenta (wave-vectors). The integrals provide the averaging over the phase of each mode, which can be motivated as follows. The modes in momentum space, all with independent phases, become arbitrarily dense in the large volume limit. Thus whenever a less dense approximation is used, each mode of the more sparse set represents the many original modes in its neighborhood, and an average over its phase represents the random phases of these neighborhood modes.

Boyer \cite{B} used (\ref{j}) to derive a correlator in the massless case. His interest lay in the electromagnetic field, and the seemingly quantum mechanical behavior of charged systems in the presence of the classical background. We avoid the explicit introduction of particles since we wish to compare classical and quantum \emph{field} theories. And our focus is on interacting fields. Boyer also emphasized the Lorentz invariance of the background configuration. This in turn is related to its possible stability in the interacting case, since fluctuations away from this configuration break the symmetry.

In summary $D_{P}$ is implicit in the description of classical evolution while $D_{\delta}$ is generated by the fluctuating background itself. In an interacting theory the classical evolution will be corrected by the fluctuating background, through the coupling between the evolving field and the background, and conversely the correlations in the background will be corrected by the interacting nature of the classical evolution. We will discuss these effects in a perturbative setting in section \ref{pe} where they give rise to a loop expansion. But we consider first a nonperturbative treatment in a simple lattice model.

\section{A lattice model\label{sa}}

The classical equations of motion govern the evolution of classical configurations in real time. To implement this evolution on a lattice there is no need to Euclideanize the theory. We will also find the lattice theory easier to implement and much faster computationally than the standard Euclidean lattice approach to quantum field theory.

For simplicity we consider a real scalar field theory in $1+1$ dimensions with latticized spatial direction. The time direction is in principle continuous, although in the end it is also discrete for computational purposes. To simplify the formulas we choose units so that $\hbar=c=a=1$ where $a$ is the lattice spacing. Then the spatial coordinate is replaced by integers $j$ and we identify $j=N$ with $j=0$, where $N$ specifies the spatial size of the lattice. The lattice Feynman propagator for a scalar with mass $m$ is (with $x\equiv(t,j)$) \begin{equation} D_{F}^{\mathrm{lat}}(x-x')=\begin{array}{c} D(x-x')\quad\textrm{for}\; t>t'\\ D(x'-x)\quad\textrm{for}\; t<t'\end{array},\end{equation} \begin{equation} D(x)=\sum_{k=-\frac{N}{2}}^{\frac{N}{2}-1}\frac{e^{-i\omega_{k}t-i2\pi kj/N}}{2N\omega_{k}}\quad\textrm{with}\;\omega_{k}=\sqrt{4\sin(k\pi/N)^{ 2}+m^{2}}.\end{equation} The real part of the Feynman propagator is the lattice cosine transform,\begin{equation} D_{\delta}^{\mathrm{lat}}(x)=\sum_{k=-\frac{N}{2}}^{\frac{N}{2}-1}\frac{ 1}{2N\omega_{k}}\cos(\omega_{k}t+\frac{2\pi kj}{N}).\end{equation} It is symmetric, $D_{\delta}^{\mathrm{lat}}(x)=D_{\delta}^{\mathrm{lat}}(-x)$, and in addition displays the spatial reflection symmetry, $D_{\delta}^{\mathrm{lat}}(t,j)=D_{\delta}^{\mathrm{lat}}(t,N-j)$. The discretized Hamiltonian density of our theory is\begin{equation} \mathcal{H}=\frac{1}{2}\dot{\phi}(t,j)^{2}+\frac{1}{2}(\phi(t,j)-\phi(t, j-1))^{2}+\frac{1}{2}m^{2}\phi(t,j)^{2}+\frac{1}{4}\lambda\phi(t,j)^{4}. \end{equation}

The fluctuating background configuration $\phi_{0}$ is a sum over modes, \begin{equation} \phi_{0}(x)=\sum_{k=-\frac{N}{2}}^{\frac{N}{2}-1}\frac{1}{\sqrt{N\omega_ {k}}}\cos(\omega_{k}t+\frac{2\pi kj}{N}+\theta_{k}).\label{a}\end{equation} The phases $\theta_{k}$ are independent parameters, each randomly distributed from $0$ to $2\pi$. These plane-waves are solutions of the free scalar field equations with discrete space and continuous time. The free Hamiltonian for this configuration for $\lambda=0$ and for any set of random phases is\begin{equation} H=\sum_{j}\mathcal{H}=\sum_{k=-\frac{N}{2}}^{\frac{N}{2}-1}\frac{1}{2}\omega_{k}\end{equation} Given that $\hbar=1$ this realizes the zero point energy spectrum of free quantum field theory.

We can now obtain the product of fields averaged over phases (using (\ref{j})), which reproduces the real part of the Feynman correlator of the lattice quantum field theory,\begin{equation} \left\langle \phi_{0}(x)\phi_{0}(x')\right\rangle _{\theta}=D_{\delta}^{\mathrm{lat}}(x-x').\label{b}\end{equation} These results are extended to $3+1$ dimensions by making suitable replacements of $N$ by volume $V$. Then in the continuum limit $D_{\delta}^{\mathrm{lat}}$ becomes $D_{\delta}$ of (\ref{k}).

We notice how the averaging over phases is associated with the finite volume approximation, which makes discrete the set of spatial momenta. Each mode of the continuum in the infinite volume case is associated with the closest discrete momentum mode in the finite volume case. Each continuum mode has its own random phase and for the single phase of the discrete momentum mode to best represent this, we average over the value of the phase of each discrete momentum mode.

The lattice theory allows us to treat the classical evolution in the presence of interactions. We can use (\ref{a}) to specify the initial condition, but evolve this forward in time according to the equations of motion with $\lambda\neq0$. Our findings indicate that the system quickly evolves toward a configuration that reflects the properties of the interacting theory, and that it does so quite efficiently. We thus measure the correlation functions at some time sufficiently well away from $t=0$. To implement the average over the phases we average over the phases that specify the initial condition. That is we generate many different spacetime dependent field configurations, each specified by the set of phases in the initial condition. We then average the contributions of all these configurations to obtain the correlation functions.

The evolution forward in time is achieved by discretizing in time and numerically solving the discretized equation of motion. The time step $a_t$, in units of the spatial lattice spacing $a$, is typically an order of magnitude smaller than unity to minimize error. We will use the leapfrog method which propagates forward $\phi(t,x)$ and $\dot{\phi}(t,x)$, given the initial values $\phi(0,j)$ and $\dot{\phi}(\frac{a_t}{2},j)$.\begin{eqnarray} \phi(t+a_t,j) & = & \phi(t,j)+a_t\dot{\phi}(t+\frac{a_t}{2},j)\nonumber \\ \dot{\phi}(t+\frac{3a_t}{2},j) & = & \dot{\phi}(t+\frac{a_t}{2},j)+a_t\left[\phi(t+a_t,j-1)-2\phi(t+a_t,j)+\phi(t+a_t,j+1)\right.\\ &  & \left.\qquad\qquad-\: m^{2}\phi(t+a_t,j)-\lambda\phi(t+a_t,j)^{3}\right]\nonumber \end{eqnarray} This method has second order accuracy and is reversible, so that energy is well conserved. The initial values are determined by (\ref{a}) except that we replace the mass $m$ by a parameter. This parameter can be made to match the physical mass extracted from the simulation, while $m$ remains the bare mass in the field equation.

For the space and time correlators at time $t_f$ we calculate \begin{eqnarray} D(0,l) & = & \left\langle \phi(t_{f},j)\phi(t_{f},[j+l]\textrm{\, mod\,}N)\right\rangle _{\theta,j},\\ D(t,0) & = & \left\langle \phi(t_{f},j)\phi(t_{f}-t,j)\right\rangle _{\theta,j},\label{e}\end{eqnarray} where we have also indicated an average over $j$. The time correlator may be a more useful object than the space correlator, since the latter falls exponentially while the former oscillates with slowly decreasing amplitude. By comparing either or both of these correlators to the corresponding free correlators with adjustable mass, we can estimate the physical mass. The extent to which the two correlators produce the same mass is an indication of how well Lorentz symmetry is respected by the dynamics.

\section{Lattice results and the gap equation}

The nonperturbative structure of the $\lambda\phi^{4}$ quantum theory in $1+1$ dimensions is well known. The main feature is a critical line in $m^{2}$-$\lambda$ space ($-\infty<m^{2}<\infty$, $\lambda>0$) occurring for negative $m^{2}$ on which a physical mass goes to zero. This line separates the symmetric phase, occurring for more positive $m^{2}$ and/or $\lambda$, and a broken symmetry phase.

In quantum field theory the first diagram in Fig.~(\ref{bbb}a) gives a correction to the bare mass of size $\delta m^{2}=3\lambda D_{F}(0)=3\lambda D_{\delta}^{\mathrm{lat}}(0)$. This one-loop effect is clearly included in the classical theory, as can be seen by averaging over the phases of two of the fields in the $\frac{\lambda}{4}\phi^{4}$ term. The same one-loop gap equation can then be used to provide a better determination of the mass in both the classical and quantum theories. We can define $m_{\mathrm{gap}}$ as the self-consistent solution to the equation\begin{equation} m_{\mathrm{gap}}^{2}=m^{2}+\left.3\lambda D_{\delta}^{\mathrm{lat}}(0)\right|_{m^{2}\rightarrow m_{\mathrm{gap}}^{2}}.\label{d}\end{equation} This one-loop gap equation sums up all the bubble graphs of the form of the first three graphs in Fig.~(\ref{bbb}a). From $m_{\mathrm{gap}}$ we can define a dimensionless coupling $g=\lambda/m_{\mathrm{gap}}^{2}$, and we expect that the gap equation gives a good representation of the full quantum theory up to effects of order $g^{2}$.

The mass and its renormalization as extracted from the classical simulation away from the critical line can be compared with the quantum result, but this comparison is carried out elsewhere \cite{C}. Here we shall concentrate on the nonperturbative physics near the critical line to study whether the classical simulation shows similar nonperturbative behavior. To focus on departures from the gap equation, we can consider increasing $g$ along a line of constant $m_{\mathrm{gap}}$. If this line crosses the critical line, then the physical mass should deviate strongly from $m_{\mathrm{gap}}$ and drop to zero at the crossing point. Recent determinations of the critical line in correspond to $g\approx10.24$ from Euclidean Monte Carlo methods \cite{A} and $g\approx9.98$ from density matrix renormalization group methods \cite{G}.

\begin{figure} \begin{center}\includegraphics[scale=0.6]{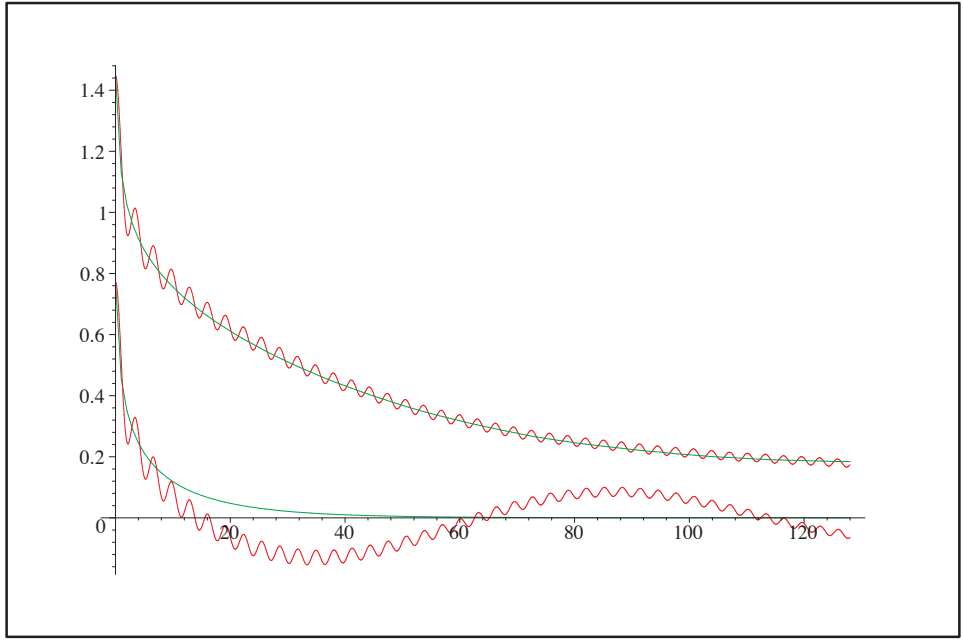}\end{center} \vspace{-3ex} \caption{\label{B} Overlapping time (wavy) and space correlators for $g=10$ ($\lambda=49.3/N^{2}$, $m^{2}=-160.3/N^{2}$, $t_{f}=0.76N$, $N=256$). For comparison the lower curves show massive free correlators given by $D^\mathrm{lat}_\delta$ with $m^2=1/N$.} \end{figure}

Since we have both the time and space correlators at our disposal, we can look for a clear signature of a vanishing physical mass. When $\omega=m$ there is nothing that distinguishes the time and space directions other than the signature of the metric, and the two correlators should coincide over time and length scales small compared to the spatial size of the lattice.\footnote{This can be seen explicitly from known results in conformal field theories upon analytic continuation of Euclidean results.} We do indeed find this phenomena. For example with $m^2$ and $\lambda$ values corresponding to $g=10$ the simulation produces the overlapping correlators pictured in Fig.~(\ref{B}), in stark contrast to the examples of free massive correlators also shown. However there is some ambiguity in this result. The point is that for this strong coupling the measured correlators have some dependence on the value of $t_f$. By adjusting $t_f$ we can obtain a similar overlap of the correlators for a range of $g$, roughly between 6 and 11. (This is along a line in $m^{2}$-$\lambda$ space with fixed $m_{\mathrm{gap}}$, with $t_f$ varying from $0.83N$ to $0.74N$, and similar results are obtained along other lines with different fixed $m_{\mathrm{gap}}$.) 

We will discuss this $t_f$ dependence in \cite{C} in the context of a slow thermalization of our system. It is significant that the quantum-like interacting configurations establish themselves so quickly, before thermalization effects become important, leaving us a window for the study of the quantum-like physics. (Effects of slow thermalization have been noted before \cite{thermstable}.) In \cite{C} we will deal with weaker couplings, where the window becomes very large compared to the time scales of interest. Our results here show that a window exists even for strong coupling, and we find in this window a signal for masslessness for a range of couplings that is consistent with the critical coupling of the quantum theory.

\begin{figure} \begin{center}\includegraphics[scale=0.6]{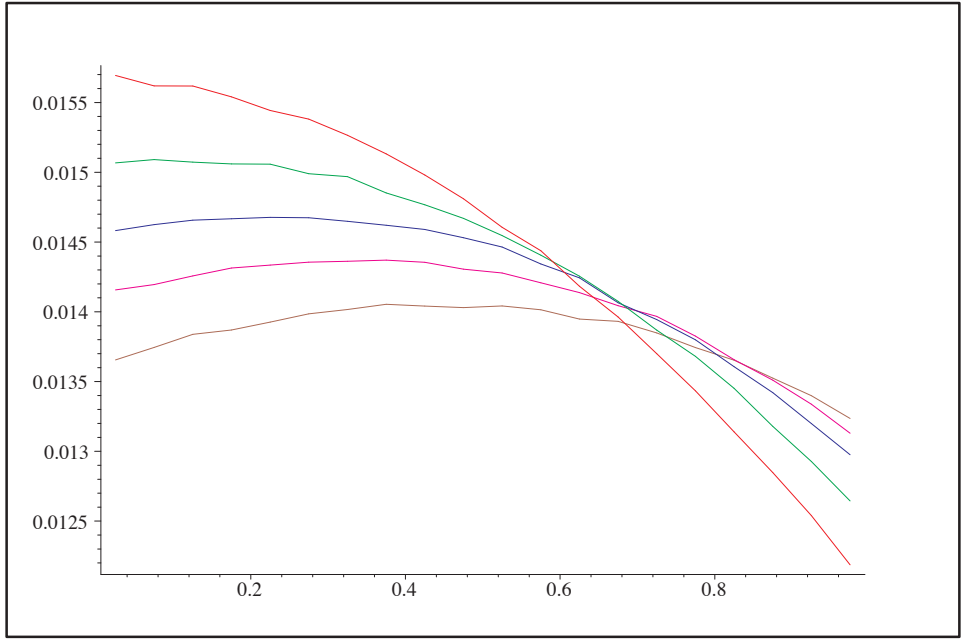}\end{center} \vspace{-3ex} \caption{\label{C}Histograms of the average value of the field for various couplings for an initial average value of 0.4. The value of the field is averaged over time slices from $t_f=N/2$ to $t_f=2N$. The values of the coupling from top to bottom are $g=5,7,10,14,20$.} \end{figure}

Another signature of a critical line is that it marks the turning on of an order parameter, a vacuum expectation value. A simple average of the classical field will not be sufficient to find this, since this will always be zero given the random initial conditions. But if the field initially starts with a nonzero average value, the question is whether there is a tendency for this average value to persist. To test for this we add a positive constant to the initial field configuration and then record the average value of the field at a later time. We bin the results to form a histogram of the final average value. A peak in this histogram at a positive value, at roughly the starting average value, would be an indication of a nonvanishing order parameter, and thus symmetry breaking. We display a sequence of such histograms in Fig.~(\ref{C}). At weak coupling the histogram is strongly peaked at zero, but the histogram flattens for larger coupling, and starts to show evidence of symmetry breaking for $g$ around 10.\footnote{For both Figs.~(\ref{B}) and (\ref{C}) the simulations were run with 10000 configurations, and $2.5/N^2$ was used as the mass-squared parameter in the initial condition.} 

We thus see that the classical theory displays critical behavior at large coupling, both in the vanishing of mass and the nonvanishing of an order parameter. These results are strikingly consistent with what is known of the quantum theory, and this certainly justifies a more detailed comparison of the two theories. The classical lattice model can be compared directly with a lattice quantum field theory, for example by precisely comparing the physical mass in the two theories. This study will be carried out in Ref.~\cite{C}, where we compare the classical theory to the quantum gap equation, at both the one- and two-loop levels. We find good agreement. That study also highlights the speed and accuracy of the classical simulation in comparison to standard Monte Carlo methods.

Another approach to the classical theory is to directly study its perturbative expansion as defined in the continuum. This will be carried out in the next few sections.

\section{Perturbative expansion without background}

First consider the scalar version of the Wheeler-Feynman classical theory without background, $\phi_{0}=0$. The perturbative expansion here will be described by tree graphs. A field theory specified by a Lagrangian $\mathcal{L}=\mathcal{L}_{0}(\phi)+\mathcal{L}_{\mathrm{int}}(\phi)+J\phi$ with interaction $\mathcal{L}_{\mathrm{int}}$ and source $J$ has classical solutions satisfying the integral equation,\begin{eqnarray} \phi(x) & = & \phi_{J}(x)+\frac{i}{\hbar}\int dyD_{P}(x-y)\frac{\delta\mathcal{L}_{\mathrm{int}}(\phi)}{\delta\phi(y)} ,\label{c}\\ \phi_{J}(x) & = & \frac{i}{\hbar}\int dyD_{P}(x-y)J(y).\end{eqnarray} $D_{P}$ is defined as the Green's function solution of (\ref{f}). The solution $\phi(x)$ is real and independent of $\hbar$; the factor $i/\hbar$ simply cancels the inverse factor in the definition of $D_{P}$.

\begin{figure} \begin{center}\includegraphics[scale=0.5]{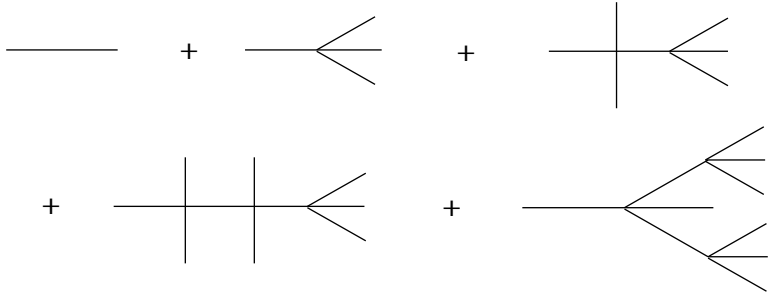}\end{center} \vspace{-3ex} \caption{Trees contributing to the solution for $\phi(x)$.\label{aaa}} \end{figure}

A solution perturbative in the coupling constant(s) in $\mathcal{L}_{\mathrm{int}}$ follows by iterating this equation. This is a solution for $\phi(x)$ in terms of $J(x)$, \begin{eqnarray} \phi(x) & = & \phi_{J}(x)+\frac{i}{\hbar}\int dyD_{P}(x-y)\mathcal{L}_{\mathrm{int}}'(\phi_{J}(y))\nonumber \\ &  & +\frac{i^{2}}{\hbar^{2}}\int dy_{1}dy_{2}D_{P}(x-y_{1})D_{P}(y_{1}-y_{2})\mathcal{L}_{\mathrm{int}}'' (\phi_{J}(y_{1}))\mathcal{L}_{\mathrm{int}}'(\phi_{J}(y_{2}))\nonumber \\ &  & +\frac{i^{3}}{\hbar^{3}}\int dy_{1}dy_{2}dy_{3}D_{P}(x-y_{1})\bigg\{ \nonumber\\ &&\qquad\hspace{2ex} D_{P}(y_{1}-y_{2})D_{P}(y_{1}-y_{3})\mathcal{L}_{\mathrm{int}}'''(\phi_{ J}(y_{1}))\mathcal{L}_{\mathrm{int}}'(\phi_{J}(y_{2}))\mathcal{L}_{\mathrm{int}}'(\phi_{J}(y_{3}))\nonumber \\ &  & \qquad +D_{P}(y_{1}-y_{2})D_{P}(y_{2}-y_{3})\mathcal{L}_{\mathrm{int}}''(\phi_{ J}(y_{1}))\mathcal{L}_{\mathrm{int}}''(\phi_{J}(y_{2}))\mathcal{L}_{\mathrm{int}}'(\phi_{J}(y_{3}))\bigg\}+\cdots \label{aa} \end{eqnarray} This expansion is shown graphically in Fig.~(\ref{aaa}) for $\lambda\phi^{4}$ theory. Each term is represented by a tree that is rooted at the point $x$, with the tips of branches representing integrals over $J$. The sum of contributions with $n-1$ tips defines a connected $n$-point function,\begin{equation} \left.\frac{\hbar\delta}{i\delta J(x_{2})}\frac{\hbar\delta}{i\delta J(x_{3})}...\frac{\hbar\delta}{i\delta J(x_{n})}\phi(x_{1})\right|_{J=0}.\end{equation} These $n$-point functions are just encoding the response of the field to a classical source.

Suppose we wanted $n$-point functions to include disconnected trees as well, so as to describe the response of a product of fields to a classical source. Then we can introduce a generating functional\begin{equation} Z[\tilde{J},J]=e^{\frac{i}{\hbar}\int dx\tilde{J}(x)\phi(x)},\end{equation} and the $n$-point function with $p$ trees is given by\begin{equation} \left.\frac{\hbar\delta}{i\delta\tilde{J}(x_{1})}...\frac{\hbar\delta}{i \delta\tilde{J}(x_{p})}\frac{\hbar\delta}{i\delta J(x_{p+1})}...\frac{\hbar\delta}{i\delta J(x_{n})}Z[\tilde{J},J]\right|_{\tilde{J}=J=0}.\label{i}\end{equation} This is a product of $p$ fields sourced by $n-p$ factors of $J$. With these definitions the tree graphs will have factors of $i$ and $\hbar$ that will reproduce the tree graphs of quantum field theory (although $D_{P}$ appears rather than the Feynman propagator). In particular the $n$-point functions are then $i^{p-n}$ times something real. The contribution at zeroth order in the coupling(s) just involves the $n$ points joined in pairs so that $p=n/2$. For general tree graphs, $1\leq p\leq n/2$.

\section{Perturbative expansion with background\label{pe}}

In this section we will see how loop graphs emerge.\footnote{This is analogous to the emergence of loops in the study of classical field theories in thermal backgrounds \cite{thermpert}.} We turn on the background field $\phi_{0}(x)$ as a solution to the free theory, $(\partial_{x}^{2}+m^{2})\phi_{0}(x)=0$. In three spatial dimensions with finite volume $V$ the background field $\phi_{0}$ takes the form\begin{equation} \phi_{0}(x)=\sum_{\mathbf{p}}\sqrt{\frac{\hbar}{\omega V}}\cos(\omega_{\mathbf{p}} t+\mathbf{p}\cdot\mathbf{x}+\theta_{\mathbf{p}}).\end{equation} Each phase $\theta_{\mathbf{p}}$ is undetermined. Now the solution to the full theory $\phi(x)$ is described by equations obtained by making the replacement $\phi_{J}\rightarrow\phi_{J}+\phi_{0}$ in (\ref{c}) and (\ref{aa}). $\phi_{0}$ is distinguished from $\phi_{J}$ since it does not vanish when $J=0$. Diagrammatically in terms of the previous tree graphs, any line that ends on $J$ can be replaced by a factor of $\phi_{0}$. We now consider how these factors are to be treated.

By averaging over the phases the generating functional and the $n$-point functions are simple extensions of what we had before.\begin{equation} Z[\tilde{J},J]=\left\langle e^{\frac{i}{\hbar}\int dx\tilde{J}(x)\phi(x)}\right\rangle _{\theta}\label{g}\end{equation} $\phi(x)$ is considered to be a function of both $J$ and $\phi_{0}$ through (\ref{aa}), with the replacement $\phi_{J}\rightarrow\phi_{J}+\phi_{0}$. The contributions to the $n$-point function with $p$-trees is written as in (\ref{i}). The averaging over phases gives nontrivial results since, although $\langle\phi_{0}\rangle_{\theta}=0$, we have\begin{eqnarray*} \langle\phi_{0}(x)\phi_{0}(y)\rangle_{\theta} & = & D_{\delta}(x-y),\\ \langle\phi_{0}(x_{1})\cdots\phi_{0}(x_{4})\rangle_{\theta} & = & D_{\delta}(x_{1}-x_{2})D_{\delta}(x_{3}-x_{4})\\&&+\;D_{\delta}(x_{1}-x_ {3})D_{\delta}(x_{2}-x_{4})+D_{\delta}(x_{1}-x_{4})D_{\delta}(x_{2}-x_{3 }),\end{eqnarray*} etc.\footnote{For the second equation we have ignored a subtlety occurring when all four phases in the average are the same. See \cite{H}.} Thus new $D_{\delta}$ lines appear that connect the tips of branches to each other, where the branches involved may be on the same tree or different trees. Trees that were formerly disconnected can become connected by $D_{\delta}$ lines. (Some trees consist only of an external point and are connected to the rest of the diagram via a $D_{\delta}$ line.) The resulting contributions to the $n$-point functions are both connected and disconnected diagrams composed of trees of $D_{P}$ lines, dressed up and interconnected with $D_{\delta}$ lines. The diagrams are considerably more complex than before, since they now include graphs with any number of loops.

The number $p$ of trees in a $n$-point function can be greater than before, now $1\leq p\leq n$. The $p$-tree contributions to an $n$-point function with $p<n$ can be derived from the quantities $\left\langle \phi(x_{1})...\phi(x_{p})\right\rangle _{\theta}$ by taking functional derivatives with respect to $J$, while the $n$-tree contribution is simply $\left\langle \phi(x_{1})...\phi(x_{n})\right\rangle _{\theta}$ in the absence of $J$. Notice that it is these $n$-tree contributions to an $n$-point that are directly probed by the correlators of the lattice model, and in particular the time and space correlators we studied correspond to the 2-tree 2-point function.

In summary the addition of a background $\phi_{0}$ generates loop effects, and a loop expansion emerges. It has the same relation between the number of the loops and the power of $\hbar$ as the loop expansion of quantum field theory. Our definition of the generating functional has ensured that the powers of $\hbar$ appearing in the propagators and vertices are the same.

\section{Rules for diagrams}

From the above discussion we see that there are two rules that characterize the set of diagrams that will be included in the calculation of an $n$-point function, following from (\ref{g}) and (\ref{i}).

\begin{enumerate} \item Each diagram has up to $n$ trees of $D_{P}$ lines, each one containing at least one external point. \item Every vertex of a diagram is on one of these trees. \end{enumerate} Clearly $D_{P}$ lines can form trees but cannot form loops by themselves. The $D_{\delta}$ have a complementary behavior, since the second rule implies that no vertex can have only $D_{\delta}$ lines attached to it. Thus while the $D_{\delta}$ lines can form loops, they cannot form trees. Correspondingly there is no diagram or sub-diagram that only has $D_{\delta}$ lines emanating from it.

We can formulate the rules in a way that more clearly describes the nature of the two types of lines by using a suggestive terminology.

\begin{itemize} \item[a)] The $D_{P}$ lines are irrotational: they cannot make up a closed loop. \item[b)] The $D_{\delta}$ lines are divergenceless: they cannot be the only type of line emerging from a graph or subgraph. \item[c)] There are no vacuum graphs, disconnected subgraphs without external lines. \end{itemize} These rules are equivalent to the previous two for the construction of $n$-point functions.

We can define amputated $n$-point functions in a way similar to quantum field theory (the LSZ reduction formula). For each coordinate $x_{i}$ of an $n$-point function we act with the operator $(i/\hbar)(\Box_{i}+m^{2})$. This vanishes when acting on $D_{\delta}$, and thus amputated graphs can only originate from graphs that have only $D_{P}$ lines as external lines. Alternatively amputated $n$-point functions can be directly obtained from the $n$-point functions defined by the generating functional\begin{equation} Z_{P}[\tilde{J},J]=\left\langle e^{\frac{i}{\hbar}\int dx\tilde{J}(x)\delta\phi(x)}\right\rangle _{\theta}\end{equation} where $\delta\phi=\phi-\phi_{0}$. These $n$-point functions have $D_{P}$ lines only as external lines, and their amputation gives the amputated $n$-point functions. This makes clear that the amputated $n$-point functions are related to the scattering of the excitations $\delta\phi$ away from the background $\phi_{0}$.

In quantum field theory the \emph{scattering amplitude} is defined as $-i$ times the sum of connected, amputated diagrams. For now we will adopt this as a tentative definition of a scattering amplitude in the classical theory, since our goal is to simply compare quantities that are defined in similar ways in the two theories. To make the comparison we need to decompose the internal lines of the quantum field theory graphs in terms of $D_{P}$ and $D_{\delta}$. Then for a given graph of the classical theory we can compare it to the corresponding decomposed piece of the quantum field theory graph.

We have defined the generating functional of the classical theory in such a way that the factors of $i$ and $\hbar$ will agree with the corresponding diagram in the quantum theory. We also note that in the classical theory the factors of $i$ are determined by the number of trees. We have already seen that the $n$-point functions are $i^{p-n}$ times something real, where $p$ is the number of trees. Thus after amputation, a contribution to a scattering amplitude is $i^{p+1}$ times something real.\footnote{Another way to see that each tree comes with a factor of $i$ is to notice that any tree contained in an amputated graph must have one more vertex than the number of $D_{P}$ lines.}

There are in addition numerical symmetry factors in the graphs of quantum field theory. We find that these symmetry factors are also reproduced by the classical theory as long as the graph makes a contribution to a scattering amplitude. We provide the details of this in Appendix A. For graphs that do not, such as disconnected graphs, or graphs with $D_{\delta}$ lines as external lines, there may be differences in their numerical factors.

Thus the only source of discrepancy in scattering amplitudes lies in those graphs that appear in quantum field theory but not in the classical theory. These are the missing graphs, the ones not satisfying the rules above. We find that the missing graphs affect the singularity structure of the momentum space integrals, since all missing graphs involve at least a pair of internal lines going on mass shell simultaneously. This is obvious for the graphs that violate rule b) above, since $D_{\delta}$ lines are on-shell by definition.\footnote{There are also missing tadpole graphs where the $D_{\delta}$ is the connecting line, but this vanishes anyway unless the connecting line is a massless scalar.} The missing graphs violating rule a) are such that a loop momentum runs around a loop composed of $D_{P}$ lines only. This loop integral vanishes in quantum field theory if the location of the poles, in terms of the loop momenta being integrated over, are distinct (since then it can be transformed into a sum of simple poles by partial fractions, each of which has a vanishing principal value integral). Only when at least one pair of poles coincide, forming a double pole, can the integral be non-vanishing. Thus at least two lines must go on-shell simultaneously.

Even with these omissions, the graphs in the classical theory still contain lines that are simultaneously on-shell. In particular a graph with $p$-trees may be cut into $p$ pieces by cutting only on-shell $D_{\delta}$ lines, with each piece containing one or more external points.

\section{2-point functions}

\begin{figure} \begin{center}\includegraphics[scale=0.5]{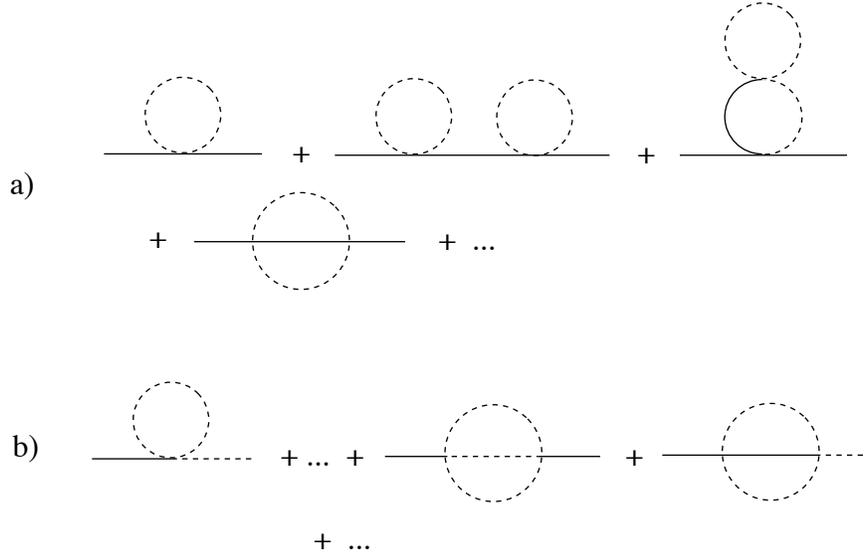}\end{center} \vspace{-3ex} \caption{Some 1-tree (a) and 2-tree (b) corrections to the 2-point function in $\lambda\phi^{4}$ theory. The solid line is $D_P$ and the dashed line is $D_{\delta}$.\label{bbb}} \end{figure}

Summing up the 1- and 2-tree contributions to the 2-point function at zeroth order in coupling gives\begin{eqnarray} &  & \left.\left(\frac{\hbar\delta}{i\delta \tilde{J}(x)}\frac{\hbar\delta}{i\delta J(x')}+\frac{\hbar\delta}{i\delta \tilde{J}(x)}\frac{\hbar\delta}{i\delta \tilde{J}(x')}\right)Z[\tilde{J},J]\right|_{\tilde{J}=J=\lambda=0}\nonumber\\ & & =\: D_{P}(x-x')+D_{\delta}(x-x')=D_{F}(x-x').\end{eqnarray} This is the Feynman propagator, emerging as the sum of two parts with distinct physical interpretations. The first is the response of the field to a source $J$, and the second is a correlation in the background field. In this section we are concerned with the perturbative loop corrections. For example for $\lambda\phi^{4}$ theory we show some 1-tree and 2-tree corrections in Fig.~(\ref{bbb}a) and (\ref{bbb}b) respectively. From the $i^{p+1}$ rule, we will refer to the 1-tree graphs as real and the 2-tree graphs as imaginary.

The 1-tree part of the 2-point function includes chains of self-energy graphs, each of which is also of the 1-tree type, that can be summed up in the usual way. The 2-tree part of the 2-point function also receives contributions from chains of self-energy graphs, such that somewhere along the chain it can be cut by cutting $D_{\delta}$ lines only. Thus there can at most be one $D_{\delta}$ line or one self-energy graph of the 2-tree type, but not both, on a bubble chain. Otherwise the bubble chain would have 3 or more trees, and this does not occur in a 2-point function.

\begin{figure} \begin{center}\includegraphics[scale=0.5]{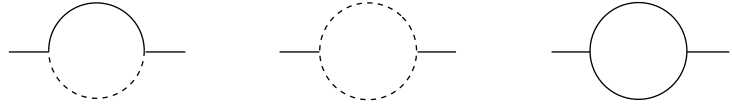}\end{center} \vspace{-3ex} \caption{The decomposed 1-loop self-energy. The last graph is missing.\label{ddd}} \end{figure}

We now show how the self-energy graphs themselves can differ from quantum field theory. This is easiest to discuss in a theory with a trilinear coupling. In particular we consider the one-loop self-energy graphs in Fig.~(\ref{ddd}). The first graph is real while the other two are imaginary. In our theory we have the real graph along with the first imaginary graph; but the second imaginary graph is missing since it violates our rule for $D_{P}$ lines. As we have described above, only when both internal lines go on-shell simultaneously, forming a double pole, does this graph contribute. In the case of a single mass $m$ in the propagators, this occurs for external momentum $p^{2}>(2m)^{2}$ or $p^{2}<0$. The two imaginary graphs in fact give contributions of the same absolute value, with the contributions adding for $p^{2}>(2m)^{2}$ and canceling for $p^{2}<0$. Thus in our theory with only the one imaginary graph contributing, the imaginary amplitude for $p^{2}>(2m)^{2}$ is smaller by half, and there is a non-vanishing contribution for $p^{2}<0$.  \begin{figure} \begin{center}\includegraphics[scale=0.5]{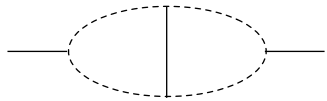}\end{center} \vspace{-3ex} \caption{An example of a missing real self-energy graph.\label{eee}} \end{figure} \begin{figure} \begin{center}\includegraphics[scale=0.5]{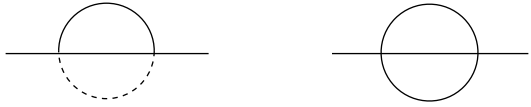}\end{center} \vspace{-3ex} \caption{Missing self-energy graphs in $\lambda\phi^{4}$ theory. The first is imaginary and the second is real.\label{fff}} \end{figure}

The $p^{2}<0$ contribution is of interest for example when the propagator appears in a $t$-channel exchange. But this does not give rise to a correction of the classical potential of theory, defined as the response of the field to a source $J$, since latter lies in the 1-tree part of the 2-point function while this particular correction contributes to the 2-tree part. To find graphs missing from the real part of the self-energy requires going to two loops. An example that violates the rule for $D_{\delta}$ lines is shown in Fig.~(\ref{eee}).

Returning to $\lambda\phi^{4}$ theory, we see a discrepancy first showing up at two loops. The two graphs that are missing (thus not appearing in Fig.~(\ref{bbb})) are shown in Fig.~(\ref{fff}), and they are real and imaginary respectively. Thus our perturbative expansion has identified these diagrams as the effects at order $\lambda^2$ that distinguish a quantum theory from the classical theory. This result is consistent with analyses involving the Schwinger-Keldysh (SK) formalism as applied to quantum field theories and their classical or high occupation number limits \cite{thermpert}. On the other hand there is a subtlety in relating these perturbative expansions to the results of our simulations. This has to do with our choice for the ensemble of classical configurations, where there is an averaging over the mode phases but not the amplitudes. This is explored further in \cite{H}.

\section{Fermions}

The Feynman propagator for fermions can be simply obtained from the one for bosons.\[ S_{F}(x-x')=(i\partial\!\!\!/_{x}+m)D_{F}(x-x').\] We again have $S_{F}=S_{P}+S_{\delta}$ with $(i\partial\!\!\!/_{x}-m)S_{P}=i\delta^{4}(x-x')$ and $(i\partial\!\!\!/_{x}-m)S_{\delta}=0$. $S_{P}$ will again arise as the classical response of the fermion field to an external source. For $S_{\delta}$, for example in $1+1$ dimensions,\footnote{We are using the representation $\gamma^{0}=\left(\begin{array}{cc} 0 & i\\ -i & 0\end{array}\right)\qquad\textrm{and}\qquad\gamma^{1}=\left(\begin{array}{cc} i & 0\\ 0 & -i\end{array}\right)$ .} we have\[ S_{\delta}(x-x')=\int\frac{dp}{(2\pi)}\frac{1}{2\omega_p}\left[\begin{array}{cc} m\cos(R)+p\sin(R) & \omega_p\sin(R)\\ -\omega_p\sin(R) & m\cos(R)-p\sin(R)\end{array}\right]\] with $R=\omega_p(t-t')+p(x-x')$. The question is whether a classical representation of $S_{\delta}$ exists as it does in the scalar case.

We shall try something closely analogous to the real mode solutions of the Klein-Gordon equation. This would be the Majorana spinor solution of the Dirac equation, which is a sum of the positive and negative frequency solutions. This results in a purely real spinor solution in the representation we are using,\[ \psi_{\theta}(x,t)=\int\frac{dp}{(2\pi)}\sqrt{2}\left(\begin{array}{c} \frac{p}{\omega_p}\cos(\omega_p t+px+\theta_p)-\frac{m}{\omega_p}\sin(\omega_p t+px+\theta_p)\\ \cos(\omega_p t+px+\theta_p)\end{array}\right).\] In the same way as before we perform the average $\left\langle \psi_{\theta}(x)\overline{\psi}_{\theta}(x')\right\rangle _{\theta}$, but in this case we do \emph{not} obtain $S_{\delta}(x-x')$. On the other hand we \emph{do} obtain $S_{\delta}(x-x')$ with the following modification,\[ \langle \psi_{\theta}(x)\overline{\psi}_{\theta+\frac{\pi}{2}}(x')\rangle _{\theta}=-\langle \psi_{\theta+\frac{\pi}{2}}(x)\overline{\psi}_{\theta}(x')\rangle _{\theta}=iS_{\delta}(x-x').\] The notation means that the relative phases are shifted before the phase averaging. The emergence of the minus sign shows how ``fermion statistics'' manifests itself in this construction. It remains to be seen whether we could use such a construction to simulate an interacting fermion theory on a lattice, as we did for the scalar, by using the Dirac equation to evolve the field and averaging over initial phases.

Alternatively we may explicitly introduce (complex) Grassman random variables $\xi_p$ to replace our previous random $c$-number variables $\theta_p$. Then a fluctuating fermionic background $\psi_0(x)$ can be obtained from quantum field operator by replacing the fermionic annihilation and creation operators by $b_p\propto\sqrt{\hbar}\xi_p$ and $b^\dag_p\propto\sqrt{\hbar}\xi^\dag_p$. $S_\delta$ is realized as $\langle \psi_0(x)\bar{\psi}_0(x')\rangle_\xi$ using the fermionic analog of averaging, i.e. \begin{eqnarray} \langle \xi_p\rangle_\xi &=& \langle \xi^\dag_p\rangle_\xi =0\\ \langle \xi_p\xi^\dag_{{q}}\rangle_\xi &=&-\langle \xi^\dag_{{q}}\xi_p\rangle_\xi =\delta_{p,{q}} \\ \langle \xi_p\xi^\dag_{{q}}\xi_{{r}} \xi^\dag_{{s}}\rangle_\xi &=&\langle \xi_p\xi^\dag_{{q}}\rangle_\xi \langle \xi_{{r}}\xi^\dag_{{s}}\rangle_\xi -\langle \xi_p\xi^\dag_{{s}}\rangle_\xi \langle \xi_{{r}}\xi^\dag_{{q}}\rangle_\xi \end{eqnarray} etc. Then as before we can iteratively solve the equation of motion and obtain perturbative expansions which contain loop graphs in the classical theory. With Grassman variables we have also been able to simulate a Yukawa theory on a lattice and verify that the fermion loop effect is included in the scalar correlator. At this order it is necessary to keep track of products of four Grassman variables, and thus it becomes increasingly difficult to go to higher orders or to obtain nonperturbative results using such a simulation.

\section{Conclusion}

We have studied a classical picture which shares many features with quantum field theory. In this picture $\hbar$ is not in any way fundamental; it merely sets the overall magnitude of the fluctuations in the classical background.  A loop expansion in powers of $\hbar$ exists in the classical theory just as it does in the quantum theory. We have developed a perturbative expansion directly from the classical equation of motion in the presence of the background, but we refer the reader to \cite{H} for a more complete comparison with quantum field theory. This type of picture may be of interest for the cosmological constant problem because it raises the notion that $\hbar$ could be a derived quantity, a quantity that may be fixed by the value the cosmological constant rather than the other way around \cite{E}.

We used a lattice to analyze the theory at the nonperturbative level, by using the full classical equation of motion to evolve configurations that only differ by the choice of phases in the initial condition. We find that the classical system evolves quickly towards configurations that incorporates features of the interacting theory. The 2-point function is easily extracted. We found a signal of the vanishing of the physical mass, the overlap of the time and space correlators, at values for the bare coupling and mass that are consistent with the known location of the critical line in $\lambda\phi^{4}$ quantum field theory in $1+1$ dimensions. We also found that the average value of the background field, the analog of a vacuum expectation value, becomes nonvanishing as the coupling increases through the critical value. This agreement with quantum field theory is surprising, since if the theories differ at the perturbative level as naively expected then there should be obvious differences of order one at strong coupling. Clearly more detailed studies are needed to understand how the classical theory is producing this strong coupling behavior.

\section*{Acknowledgments} We thank J.~Giedt, R.~Koniuk and T.~Yavin for discussions. This work was supported in part by the National Science and Engineering Research Council of Canada.

\appendix \section{Symmetry factors}

We explain why the symmetry factor of a graph contributing to a scattering amplitude is the same as in quantum field theory. The perturbative expansion of generating functional for the classical theory is \begin{eqnarray} &&\hspace{-6ex} Z[\tilde{J},J] \nonumber\\ &=& \left\langle e^{\frac{i}{\hbar}\int dx\tilde{J}(x)\phi(x)}\right\rangle_{\theta} \nonumber\\ &=&\left\langle\exp \frac{i}{\hbar}\int dx\tilde{J}(x)\Bigg\{\Phi_0(x) +\frac{i}{\hbar}\int dy D_P(x-y){\cal L'}_{int}(\Phi_0(y))\right. \nonumber\\&&\qquad\left. -\frac{1}{\hbar^2}\int dy_1dy_2 D_P(x-y_1)D_P(y_1-y_2) {\cal L''}_{int}(\Phi_0(y_1)){\cal L'}_{int}(\Phi_0(y_2)) +\cdots\Bigg\}\right\rangle_\theta \\ &=&\left\langle e^{\frac{i}{\hbar}\int dx\tilde{J}\Phi_0} \right\rangle_\theta -\frac{1}{\hbar^2}\int dxdy\tilde{J}(x)D_P(x-y) \left\langle e^{\frac{i}{\hbar}\int dx'\tilde{J}\Phi_0} {\cal L'}_{int}(\Phi_0(y)) \right\rangle_\theta \nonumber\\&& -\frac{i}{\hbar^3}\int dxdy_1dy_2\tilde{J}(x)D_P(x-y_1) D_P(y_1-y_2)\left\langle e^{\frac{i}{\hbar} \int dx'\tilde{J}\Phi_0}{\cal L''}_{int}(\Phi_0(y_1)) {\cal L'}_{int}(\Phi_0(y_2))\right\rangle_\theta \nonumber\\ &&+\frac{1}{2\hbar^4}\int dx_1dx_2dy_1dy_2\tilde{J}(x_1) D_P(x_1-y_1)\tilde{J}(x_2)D_P(x_2-y_2) \left\langle e^{\frac{i}{\hbar}\int dx'\tilde{J}\Phi_0} {\cal L'}_{int}(\Phi_0(y_1)){\cal L'}_{int}(\Phi_0(y_2)) \right\rangle_\theta \nonumber\\&&+\cdots , \end{eqnarray} where $\Phi_0=\phi_J+\phi_0$. When one expands the generating functional in quantum field theory after canceling out the disconnected vacuum graphs, one gets the same expansion except that the propagator is replaced by $D_F$ and the contraction inside a bra-ket is done by $D_F$ rather than $D_{\delta}$.

From this similarity and since there is no $J$ derivative involved to obtain a $n$-tree graph of $n$-point function (thus $\Phi_0=\phi_0$ and as well the $n$ external points are undistinguishable), the symmetry factor for a $n$-tree graph of $n$-point function is seen to be same as that of the topologically same $n$-point Feynman graph in quantum field theory.

Now consider a $n$-tree graph containing the factors $D_\delta(x_{p+1}-y_{p+1}), \cdots, D_\delta(x_n-y_n)$, where $y_{p+1}\cdots y_n$ are internal points or other (i.e. $x_1$ to $x_{p}$) external points. Such a graph does not contribute to the scattering amplitude. Since $\phi_J$ appears in the combination $\phi_J+\phi_0$, a $p$-tree graph of the $n$-point function is obtained by replacing the previous factors by $D_P(x_{p+1}-y_{p+1}), \cdots, D_P(x_n-y_n)$ respectively. Then if all the external points are connected to internal points by $D_P$ lines the resulting $p$-tree graph can contribute to a scattering amplitude. Since the original $n$-tree graph has the correct symmetry factor, this $p$-tree graph does too.

Other $p$-tree graphs that do not contribute to a scattering amplitude can in general have symmetry factors that differ from quantum field theory. This can be verified by example.

\end{document}